\documentclass[12pt]{article}
\usepackage{times}
\begin{document}
\pagestyle{plain}
\hsize = 6.1 in 				
\vsize = 8.5 in		   
\hoffset = -0.5 in
\voffset = -0.5 in
\baselineskip = 0.29 in

\def\vn{\mbox{\boldmath$n$}}
\def\vx{\mbox{\boldmath$x$}}

\title{Stochastic Physics, Complex Systems and Biology\footnote{The 1st Gordon Research Conference on ``Stochastic Physics in Biology'',
chaired by K.A. Dill, was held on January 23-28, 2011, in Ventura, CA.}}
 
\author{Hong Qian\\[12pt]
Department of Applied Mathematics\\
University of Washington\\ 
Seattle, WA 98195, U.S.A.
}

\maketitle

\begin{abstract}
In complex systems, the interplay between nonlinear and 
stochastic dynamics, e.g., J. Monod's necessity and
chance, gives rise to an evolutionary process 
in Darwinian sense, in terms of discrete jumps among
attractors, with punctuated equilibrium, spontaneous
random ``mutations'' and ``adaptations''.  On an 
evlutionary time scale it produces sustainable diversity
among individuals in a homogeneous population rather
than convergence as usually predicted by a deterministic 
dynamics.  The emergent discrete states in such a 
system,  i.e., attractors, have natural  
robustness against both internal and external perturbations.
Phenotypic states of a biological cell, a mesoscopic 
nonlinear stochastic open biochemical system, could be understood 
through such a perspective.        
\end{abstract}

	Biological systems and processes are complex.  One of 
the hallmarks of complex behavior is uncertainties, either in 
the causes of an occurred event, or in predicting its 
future \cite{mackey_rmp,ge_dill}.   This ``feel'' of complexity 
is intimately related to the following issue \cite{jjh_94}: 
When a system consists of only a few degrees of
freedom, say $x_1$, $x_2$ and $x_3$, a complete
description of the ``trajectory'' of $(x_1,x_2,x_3)(t)$ for 
all $t$ consitutes a full understanding of the system.
However, when a system has a million of degrees of 
freedom, $\vx(t)=\{x_i(t)|1\le i\le 10^6\}$, a complete 
description of the $\vx(t)$ is not informative
at all!  One needs to find an particular ``angle'' to 
synthesize the large amount of data, or a ``pattern'' to
obtain a summary.  In classical physics of inanimated matters
with relatively homogeneous individuals, 
this is accomplished by introducing the concept of {\em distribution} together with macroscopic thermodynamic quantities, giving rise to the discipline of statistical thermodynamics. 
In modern cellular biology, this is known as ``data interpretation with respect to biological functions'': Usually a narrative in addition to the data is required \cite{knight}. 
	
	The foregoing brief discussion points to a key
departure from the classical physics of Newton and Laplace 
\cite{prigogine}: 
A rational choice of mathematical descriptions of biological
systems and processes requires a probabilistic view of the 
dynamics, which provides both individual-based and 
distribution-based perspectives.  
Studying system dynamics in terms of stochastics, 
either due to intrinsic uncertainties, lack of
full knowledge, or due to a need for organizing large amount of 
data, is the basis of what we call {\em stochastic physics}.

\section*{What is Stochastic Physics}

	Modern sciences emphasize quantitative representation
of experimental observations, widely known as {\em mathematical
modeling}.  Along this line, there are two types of 
modeling: the {\em data-driven} and the {\em mechanism
based} models.  In the history of physics, Kepler's model
(laws) was the most celebrated example of the former, while
Newton's theory of universal gravity, which ``explains''
Kepler's results, is the canonical
example of a mechanism.  In fact, the very term {\em mechanism}
was derived from the word {\em mechanics}.  In biology,
Mendel's model (law) was the former, and Hardy-Weinberg's theory was 
the latter.
The difference between the example in physics and the 
example in biology is that the latter has to take
into account of uncertainties.  Data-driven modeling incorporating
uncertainties gives rise to the entire field of
statistics - and bioinformatics and financial engineering
are two most active branches of studies in recent years.

This leads straight to the question ``where is the mechanism
based modeling with uncertainty''?  Stochastic physics is precisely
the answer to this calling.  In sociology and economics,
this type of modeling is called {\em agent-based}, and 
in finance it is called {\em behavior finance}.

In applied mathematics, statistics is associted 
with {\em data-driven modeling} and stochastic
process is associted with {\em population distribution
based mechanistic modeling}.  In physics, statistical
physics has traditionally dealt with more on 
state of matters in equilibrium rather than 
dynamics of open, driven systems. Nevertheless, it is
a shining example of successful stochastic modeling.

\section*{Nonlinear Physics and Stochastic Physics}

	Stochastic physics shares many
of the concepts and concerns of the nonlinear
physics that has gone before it: They both are focused 
on dynamics of a system \cite{haken_book}. Technically,
for nonlinear systems exhibiting chaotic dynamics,
a characterization based on distribution turns out 
to be more appropriate \cite{lasota}.  Data analyses
of chaotic signals also constantly employ methods
from statistics \cite{abarbanel_rmp,tong_book}.  

	Stochastic dynamics in linear systems and nonlinear
systems are fundamentally different \cite{qian_pccp,qian_nonl}.
The former can be essentially represented by a 
Gaussian process, which was extensively studied by 
eminent physicsts like Uhlenbeck, Chandrasekhar, and
Onsager \cite{wax_book,onsager_53,fox}. But stochastic dynamics 
{\it per se} is not the reason for complex behavior.  
A Gaussian process has certain unpredictability, nevertheless
the ultimate fate of the dynamics is all the same: It 
fluctuates around its mean value.

	However, when one faces a strongly nonlinear system 
with stochasticity, one has to talk about {\em evolution}, 
evolution process in Darwin's sense with punctuated equilibrium 
and spontaneous random ``mutations'' and ``adaptations''.  
This is one of 
the profound insights derived from the studies of nonlinear
stochastic systems:  The fluctuations in a nonlinear
system with multiple attractors make rare events,
something with infinitesimal probability from a 
determinsitic stand point, an sure occurance with probability
one in an ``evolutionary'' time scale \cite{ge_qian_11,qian_ge_mcb}. 
This picture fits J. Monod's notion of chance and necessity
\cite{monod_book,haken_book}.
Furthermore, when encountering external environmental changes, 
nonlinear multi-stable systems exhibit adaptation by 
enhanced rate of transition into the ``favored attractors''; 
and ultimately exhibit ``rupture'' - the nonlinear catastrophe 
scenario in the presence of stochasticity \cite{shapiro_qian_bpc_97}.  

	Newton-Laplace's dynamics gives us a sense of
convergence. For strongly nonlinear stochastic dynamics,
the validity of the converging dynamics is only on a
rather limited time scale.  In an evolutionary time scale,
divergent dynamics emerges.  This, we believe, is a
philosophical implication derived from stochastic 
physics \cite{prigogine}.

\section*{Stochastic Physics and Quantitative Biology}

Physics and computer science (CS) are two cornerstones
of modern, engineering world.  Therefore, it is
not surprising that they support the most quantitative 
aspects of biology. Yet, upon a more careful reflection, one
realizes that thinkings in both physics and CS are in odd with
that of biologists: Physics considers systems that can be described with a few variables, known as ``information
poor'' according to J.J. Hopfield \cite{jjh_94}, and CS, while 
deals with much more complex problems, nevertheless in terms
of perfect logics with almost infinite precision.  Biological
systems are information rich, and biological processes are 
not about percision or optimal, but rather about functional and
survival.

	The studies of biological cells, the universal building block of living organisms, also have two foundations that echoed physics and CS: biochemistry and genomics.  Biochemistry is founded on the tradition of physics, via the investigations of macromolecular structures and dynamics and biochemical reactions, while genomics heavily utilizes concepts and methods from CS, i.e. coding,
information, discrete mathematics, leading to the emergence of 
bioinformatics in recent years.  The heavy influences of physics and CS in biological thinking, in fact in all 21-century modern thinkings, 
is unmistakable.  Nowadays, even the studies of biochemical reaction systems are usually about their information logic flow. Known as signal transduction, it provides a 
clear link between biochemistry within a cell, to perceived function.  However, one often forgets that information is only 
an abstract term; its physical bases have to be either energy or material.  In cell biology, they are represented by the structure and states of macromolecules.  The information logic flow aspects of 
biochemical reaction is our ``models'' and ``interpretations'' of a 
biological organism based on our understanding of its engineering functions!  It is a ``narrative'' cell biologists provide to understand a complex reality \cite{petermoore}.

	This reveals an important gap in the current dominant thinking of cell biology: the link between the physics of molecules, the chemistry of reactions, and the information logic flow they represents. It is widely recognized that investigations into this
link require statistical physics and molecular thermodynamics in {\em small systems} with {\em dynamics} \cite{phillips_06,bustamante_05,qian_jbp_12}. Filling this gap has been called for as {\em the systems biology of cells} \cite{palsson_04}.
Though yet to be proven, it is not difficult to see that the 
stochastic physics approach as described above has the potential 
to be a powerful, quantitative language of cellular dynamics and 
other biological systems \cite{qian_arbp_12}.

	The stochastic physics approach to biology relies more
on mechanistic understanding of biological systems and processes
than on high-throughput large data sets.  It is a powerful tool to generate
working hypotheses in a rigorous way.   In current biological
research, one often states that ``we like to know how it works''.
However, a scientifically more sound statement should be 
``we like to know whether it works in {\em this} way?''.  
This goes back to the hypothesis-driven research
with strong inference \cite{beard_09}.  Taking uncertainties
into account, stochastic modeling is based on 
one's mechanistic understanding,
and relies on mathematical deduction to generate precise
hypothesis. It will be an indispensible tool in 
biological research on par with data-driven bioinformatics.

\section*{Cellular Biology and Theory of Evolution}

Based on the Modern Synthesis of Darwin's theory of 
evolution, the current population genetics and 
genomics \cite{koonin} attribute the molecular basis 
of biological variations to different DNA sequences, 
which is inheritable through Mendelian genetics and
Watson-Crick base-pairing mechanism.  Biochemistry, 
however, has been always considered as merely a 
deterministic mechanics that executes the instructions 
coded in the DNA \cite{alberts_98}.

	Recent laboratory measurements on {\em stochastic 
gene expression} in single cells with single-molecule 
sensitivity, however, has broken the genomic monoplay 
of biological variations 
\cite{siggia_science_02,xie_nature_06}.  Stochasticity 
has been increasingly recognized as a key aspect of 
cellular molecular biology. 
In terms of Darwin's evolution, Kirschner and Gerhart 
have maintained that the essential role of cellular and organismal 
biology is to provide phenotypic variations with plausible
molecular mechanisms that bridge genomes and lives \cite{K_and_G_book}.

	The tenants of stochasitc physics fit this
perspective. In particular, the mathematical theory of 
stochastic processes has revealed a rich thermodynamic
structure in any stochastic dynamics based on Markov
formalism \cite{ge_qian_10}.  The thermodynamic theory
clearly distinguishes a closed stochastic
system which reaches an equilibrium distribution
with detailed balance, and an open, driven stochastic
system which reaches a nonequilibrium steady state
\cite{zqq,gqq,jqq_book,bertalanffy}.  It has been firmly established that
the latter corresponds to precisely cellular
biochemical systems upon which continuous chemical
driving forces are applied.  The conversion of chemical 
energy into heat in isothermal cellular systems 
can be characterized by entropy production rate
\cite{qian_arpc_07,wangjin_08}.

	The external energy supply, as the ``environment 
condition'' for an open system, is the thermodynamic 
necessity for self-organization and complex behavior \cite{qian_arpc_07}.  Thermodynamics, however, can
only tell what is possible and what is not; but
it does not tell what is feasible and what is 
the mechanism.  For the latter, detailed
``molecular mechanisms'' have to be developed.
There is clearly a dichotomy between the
nature vs. nurture for the function of a 
biochemical system.  A stochastic description
of dynamics provides a unique tool to understand 
the occurrence of sequential events, 
i.e., kinetics, in terms of the ``most probable 
path'' \cite{wang_mpp,wang_pnas_11,ge_qian_ijmpb}.

	There is a growing interest in understanding
cell differentiation including stem cell 
differentiation and reprograming, isogenetic 
variations, and even cancer carcinogenesis from 
an evolutionary perspective at the cellular level 
\cite{weinberg,aoping_07,wang_pnas_11}.
The mathematical theory of evolution and
population genetics has long been based on 
stochastic processes \cite{ewens_book,aoping_05,aoping_08}. 
Therefore, the stochastic physics approach to
cellular biochemical dynamics provides a 
natural unifying framework to further this 
exciting new frontier of biological science.

	A stochastic physics based quantitative understanding
of cellular biology, in return, will provide a  
paradigm for studying other complex systems 
\cite{qian_nonl,zqq,qian_decomp}.

\bibliographystyle{plain}
\bibliography{spb}

\begin{thebibliography}{10}

\bibitem{abarbanel_rmp}
H~D~I Abarbanel.
\newblock The analysis of observed chaotic data in physical systems.
\newblock {\em Rev Mod Phys}, 65:1331--1392, 1993.

\bibitem{alberts_98}
B~Alberts.
\newblock The cell as a collection of protein machines: Preparing the next
  generation of molecular biologists.
\newblock {\em Cell}, 92:291--294, 1998.

\bibitem{aoping_05}
P~Ao.
\newblock Laws in {D}arwinian evolutionary theory.
\newblock {\em Phys Life Rev}, 2:117--156, 2005.

\bibitem{aoping_08}
P~Ao.
\newblock Emerging of stochastic dynamical equalities and steady state
  thermodynamics from {D}arwinian dynamics.
\newblock {\em Comm Theoret Phys}, 49:1073--1090, 2008.

\bibitem{aoping_07}
P~Ao, D~Galas, L~Hood, and Zhu X-M.
\newblock Cancer as robust intrinsic state of endogenous molecular-cellular
  network shaped by evolution.
\newblock {\em Med Hypoth}, 70:678--684, 2007.

\bibitem{beard_09}
D~A Beard.
\newblock Strong inference for systems biology.
\newblock {\em PLoS Comp Biol}, 5:e1000459, 2009.

\bibitem{bustamante_05}
C~Bustamante, J~Liphardt, and F~Ritort.
\newblock The nonequilibrium thermodynamics of small systems.
\newblock {\em Phys Today}, 58(July):43--48, 2005.

\bibitem{xie_nature_06}
L~Cai, N~Friedman, and X~Xie.
\newblock Stochastic protein expression in individual cells at the single
  molecule level.
\newblock {\em Nature}, 440:358--362, 2006.

\bibitem{shapiro_qian_bpc_97}
Shapiro~B E and H~Qian.
\newblock A quantitative analysis of single protein-ligand complex separation
  with the atomic force microscope.
\newblock {\em Biophys Chem}, 67:211--219, 1997.

\bibitem{siggia_science_02}
M~B Elowitz, A~J Levine, Siggia~E D, and P~S Swain.
\newblock Stochastic gene expression in a single cell.
\newblock {\em Science}, 297:1183--1186, 2002.

\bibitem{ewens_book}
W~J Ewens.
\newblock {\em Mathematical Population Genetics I. Theoretical Introduction}.
\newblock Springer, New York, 2004.

\bibitem{fox}
R~F Fox.
\newblock Gaussian stochastic processes in physics.
\newblock {\em Phys. Rep.}, 48:179--283, 1978.

\bibitem{ge_dill}
H~Ge, S~Press\'{e}, K~Ghosh, and K~A Dill.
\newblock Markov processes follow from the principle of maximum caliber.
\newblock {\em J Chem Phys}, 136:064108, 2012.

\bibitem{ge_qian_10}
H~Ge and H~Qian.
\newblock The physical origins of entropy production, free energy dissipation
  and their mathematical representations.
\newblock {\em Phys Rev E}, 81:051133, 2010.

\bibitem{ge_qian_11}
H~Ge and H~Qian.
\newblock Nonequilibrium phase transition in mesoscoipic biochemical systems:
  From stochastic to nonlinear dynamics and beyond.
\newblock {\em J R Soc Interf}, 8:107--116, 2011.

\bibitem{ge_qian_ijmpb}
H~Ge and H~Qian.
\newblock Analytical mechanics in stochastic dynamics: {M}ost probable path,
  large-deviation rate function and {H}amilton-{J}acobi equation.
\newblock {\em Int J Mod Phys B}, 26:1230012, 2012.

\bibitem{gqq}
H~Ge, M~Qian, and H~Qian.
\newblock Stochastic theory of nonequilibrium steady states ({P}art {II}):
  Applications in chemical biophysics.
\newblock {\em Phys Rep}, 510:87--118, 2012.

\bibitem{haken_book}
H~Haken.
\newblock {\em Synergetics, an Introduction: Nonequilibrium Phase Transitions
  and Self-Organization in Physics, Chemistry, and Biology}.
\newblock Springer-Verlag, New York, 3rd rev. enl. edition, 1983.

\bibitem{weinberg}
D~Hanahan and R~A Weinberg.
\newblock The hallmarks of cancer.
\newblock {\em Cell}, 100:57--70, 2000.

\bibitem{jjh_94}
J~J Hopfield.
\newblock Physics, computation, and why biology looks so different?
\newblock {\em J Theoret Biol}, 171:53--60, 1994.

\bibitem{jqq_book}
D-Q Jiang, M~Qian, and M-P Qian.
\newblock {\em Mathematical Theory of Nonequilibrium Steady States On the
  frontier of probability and dynamical systems, LNM Vol. 1833}.
\newblock Springer-Verlag, Berlin, 2004.

\bibitem{K_and_G_book}
M~W Kirschner and J~C Gerhart.
\newblock {\em The Plausibility of Life: Resolving {D}arwin's Dilemma}.
\newblock Yale Univ. Press, New Haven, CT, 2005.

\bibitem{knight}
J~Knight.
\newblock Bridging the culture gap.
\newblock {\em Nature}, 419:244--246, 2002.

\bibitem{koonin}
E~V Koonin.
\newblock Darwinian evolution in the light of genomics.
\newblock {\em Nucleic Acids Res}, 37:1011--1034, 2009.

\bibitem{lasota}
A~Lasota and M~C Mackey.
\newblock {\em Chaos, Fractals and Noise: Stochastic Aspects of Dynamics}.
\newblock Springer-Verlag, New York, 1994.

\bibitem{mackey_rmp}
M~C Mackey.
\newblock The dynamic origin of increasing entropy.
\newblock {\em Rev Mod Phys}, 61:981--1016, 1989.

\bibitem{monod_book}
J~Monod.
\newblock {\em Chance and Necessity: An Essay on the Natural Philosophy of
  Modern Biology}.
\newblock Vintage Books, New York, 1972.

\bibitem{petermoore}
P~B Moore.
\newblock How should we think about the ribosome?
\newblock {\em Ann Rev Biophys}, 41:1--19, 2012.

\bibitem{onsager_53}
L~Onsager and S~Machlup.
\newblock Fluctuations and irreversible processes.
\newblock {\em Phys Rev}, 91:1505--1512, 1953.

\bibitem{phillips_06}
R~Phillips and S~R Quake.
\newblock The biological frontier of physics.
\newblock {\em Physics Today}, 59(May):38--43, 2006.

\bibitem{prigogine}
I~Prigogine and I~Stengers.
\newblock {\em Order Out of Chaos: Man's New Dialogue with Nature}.
\newblock New Sci. Lib. Shambhala, Boulder, CO, 1984.

\bibitem{qian_arpc_07}
H~Qian.
\newblock Phosphorylation energy hypothesis: Open chemical systems and their
  biological functions.
\newblock {\em Ann Rev Phys Chem}, 58:113--142, 2007.

\bibitem{qian_nonl}
H~Qian.
\newblock Nonlinear stochastic dynamics of mesoscopic homogeneous biochemical
  reactions systems - an analytical theory.
\newblock {\em IoP Nonlinearity}, 24:R19--R49, 2011.

\bibitem{qian_arbp_12}
H~Qian.
\newblock Cooperativity in cellular biochemical processes: Noise-enhanced
  sensitivity, fluctuating enzyme, bistability with nonlinear feedback, and
  other mechanisms for sigmoidal responses.
\newblock {\em Ann Rev Biophys}, 41:179--204, 2012.

\bibitem{qian_decomp}
H~Qian.
\newblock A decomposition of irreversible diffusion processes without detailed
  balance.
\newblock {\em arXiv.org/abs/1204.6496}, 2012.

\bibitem{qian_jbp_12}
H~Qian.
\newblock Hill's small systems nanothermodynamics: A simple macromolecular
  partition problem with a statistical perspective.
\newblock {\em J Biol Phys}, 38:201--207, 2012.

\bibitem{qian_ge_mcb}
H~Qian and H~Ge.
\newblock Mesoscopic biochemical basis of isogenetic inheritance and
  canalization: {S}tochasticity, nonlinearity, and emergent landscape.
\newblock {\em MCB: Mol Cellu Biomech}, 9:1--30, 2012.

\bibitem{qian_pccp}
H~Qian, P-Z Shi, and J~Xing.
\newblock Stochastic bifurcation, slow fluctuations, and bistability as an
  origin of biochemical complexity.
\newblock {\em Phys Chem Chem Phys}, 11:4861--4870, 2009.

\bibitem{tong_book}
H~Tong.
\newblock {\em Non-Linear Time Series: A Dynamical System Approach}.
\newblock Oxford, UK, 1993.

\bibitem{bertalanffy}
L~von Bertalanffy.
\newblock The theory of open systems in physics and biology.
\newblock {\em Science}, 111:23--29, 1950.

\bibitem{wangjin_08}
J~Wang, L~Xu, and E~K Wang.
\newblock Potential landscape and flux framework of non-equilibrium networks:
  Robustness, dissipation and coherence of biochemical oscillations.
\newblock {\em Proc Natl Acad Sci USA}, 105:12271--12276, 2008.

\bibitem{wang_mpp}
J~Wang, K~Zhang, and E~K Wang.
\newblock Kinetic paths, time scale, and underlying landscapes: A path integral
  framework to study global natures of nonequilibrium systems and networks.
\newblock {\em J Chem Phys}, 133:125103, 2010.

\bibitem{wang_pnas_11}
J~Wang, K~Zhang, L~Xu, and E~K Wang.
\newblock Quantifying the {W}addington landscape and biological paths for
  development and differentiation.
\newblock {\em Proc Natl Acad Sci USA}, 108:8257--8262, 2011.

\bibitem{wax_book}
N~Wax.
\newblock {\em Selected Papers on Noise and Stochastic Processes}.
\newblock Dover, New York, 1954.

\bibitem{palsson_04}
H~V Westerhoff and B~{\O} Palsson.
\newblock The evolution of molecular biology into systems biology.
\newblock {\em Nature Biotech}, 22:1249--1252, 2004.

\bibitem{zqq}
X-J Zhang, H~Qian, and M.~Qian.
\newblock Stochastic theory of nonequilibrium steady states and its
  applications ({P}art {I}).
\newblock {\em Phys Rep}, 510:1--86, 2012.

\end{thebibliography}

\end{document}